\documentclass[showpacs,10pt,twocolumn,prb]{revtex4-1}

\usepackage{amsmath}
\usepackage{amssymb}
\usepackage{graphicx}
\usepackage{amssymb}
\usepackage{graphics}
\usepackage{epsfig}
\usepackage{color}

\begin{document}

\title{Multiband electronic transport in  $\alpha$-Yb$_{1-x}$Sr$_{x}$AlB$_{4}$ [x = 0, 0.19(3)] single crystals}
\author{Hyejin Ryu$^{1,2,\dag}$, Milinda Abeykoon$^{1}$, Emil Bozin$^{1}$, Yosuke Matsumoto$^{3}$, S. Nakatsuji$^{3}$ and C. Petrovic$^{1,2,3,\ddag}$}
\affiliation{$^{1}$Condensed Matter Physics and Materials Science Department, Brookhaven National Laboratory, Upton, New York 11973, USA\\
$^{2}$Department of Physics and Astronomy, Stony Brook University, Stony Brook, New York 11794-3800, USA\\
$^{3}$Institute for Solid State Physics, University of Tokyo, Kashiwa, Chiba 277-8581, Japan}
\begin{abstract}
We report on the evidence for the multiband electronic transport in $\alpha$-YbAlB$_{4}$ and $\alpha$-Yb$_{0.81(2)}$Sr$_{0.19(3)}$AlB$_{4}$. Multiband transport reveals itself below 10 K in both compounds via Hall effect measurements, whereas anisotropic magnetic ground state sets in below 3 K in $\alpha$-Yb$_{0.81(2)}$Sr$_{0.19(3)}$AlB$_{4}$. Our results show that Sr$^{2+}$ substitution enhances conductivity, but does not change the quasiparticle mass of bands induced by heavy fermion hybridization.
\end{abstract}

\maketitle

\section{Introduction}
Heavy-fermion systems have attracted much attention due to the large magnetic enhancement of quasiparticle mass, non-Fermi-liquid (NFL) behavior and unconventional
superconductivity that is believed to arise without the mediating role of phonons \cite{Lonzarich}. The breakdown of Landau Fermi liquid framework at the magnetic boundary is often intimately connected with Quantum Critical Point (QCP) \cite{Stewart,Basov,Si}. Therefore spin-fluctuation mediated pairing may be relevant not only to heavy fermion superconductors (HFSC) but also to a wide class of materials of current interest \cite{Scalapino}.  Furthermore, the charge (valence) fluctuations were also suggested to play a role in superconductivity of some HFSC at ambient or high pressures such as CeCu$_{2}$(Si$_{1-x}$Ge$_{x}$)$_{2}$ and CeRhIn$_{5}$ \cite{Watanabe1,Yuan1,Yuan2,Watanabe2}.

Quantum valence criticality was proposed to be the explanation of the unconventional critical phenomena observed in $\beta$-YbAlB$_{4}$ \cite{Watanabe1}. The $\beta$-YbAlB$_{4}$ crystallizes in the orthorhombic structure, space group \textit{Cmmm} \cite{Macaluso}, and is a first Yb based analog of Ce HFSC that shows NFL behavior associated with QCP in the normal state ($\rho$$\sim$T$^{1.5}$ and $C/T \sim lnT$ as $T\rightarrow0$) \cite{Nakatsuji1}. At nearly all ranges of thermodynamic parameters it coexists with $\alpha$-YbAlB$_{4}$ which crystallizes in the orthorombic structure, space group \textit{Pbam}, and is a well defined Fermi liquid ($\rho$$\sim$T$^{2}$ and $C/T$$\sim$$const.$ as $T\rightarrow0$) \cite{Fisk,Matsumoto1}. Both crystal structures contain Yb and Al atoms sandwiched between boron layers, with somewhat different motifs for $\alpha$ and $\beta$ phase. Despite the similarities in their crystal structure, the ground states of the two polymorphs of YbAlB$_{4}$ are rather different. Quantum valence criticallity in $\beta$-YbAlB$_{4}$ \cite{Watanabe1} is at odds with experimental observation that both polymorphs exhibit similar magnetic ground state and strong valence fluctuations with Yb valence estimated to be 2.73 and 2.75 for $\alpha$ and $\beta$ polymorphs respectively \cite{Okawa,TerashimaT}. Hence, it is of interest to investigate charge transport in $\alpha$-YbAlB$_{4}$ and its similarities or differences with $\beta$-YbAlB$_{4}$.

Here, we report on the multiband electronic transport in $\alpha$-YbAlB$_{4}$ and $\alpha$-Yb$_{0.81(2)}$Sr$_{0.19(3)}$AlB$_{4}$. Upon 19(3) \% Sr substitution on $\alpha -$YbAlB$_{4}$,
antiferromagnetic state is induced at $T_{N}$ = 3 K. As expected for a Yb-based compound \cite{Joe,Winkelmann}, contraction of the unit cell promotes the magnetic state in $4f^{13}$ by creating Kondo-hole and increasing the carrier density.

\section{Experiment}
Single crystals of $\alpha -$YbAlB$_{4}$ and $\alpha -$Yb$_{0.81(2)}$Sr$_{0.19(3)}$AlB$_{4}$ were prepared by self-flux method. Yb$_{1-x}$Sr$_{x}$, B and Al were mixed in
1:1:100 stoichiometric ratio, heated in alumina crucibles under an Ar atmosphere up to 1673 K, cooled down to 1073 K at 3.33 K/h and then cooled naturally in the furnace to the room temperature.
Excess Al flux was removed by subsequent centrifugation at 1073 K and by etching in NaOH solution. Plate-like crystals up to 1$\times $1$\times $0.5 mm$^{3}$ were found. Finely pulverized samples were filled into 1 mm
diameter Kapton capillaries for x-ray diffraction (XRD) measurements using cyllindrical transmission geometry at X7B beamline of the National Synchrotron Light Source (NSLS)
at the Brookhaven National Laboratory. Data for both samples were collected at room temperature utilizing a 0.5 mm$^{2}$ monochromatic beam with energy of ~38 keV (0.3196 ${\AA}$) and Perkin Elmer two-dimensional (2D) image plate detector mounted orthogonal to the beam path 376.4  mm away from the sample.  High quality data were collected up to $Q = 4 \pi Sin(\theta)/\lambda = 17\AA^{-1}$. Chemical compositions of crystals were obtained by energy-dispersive X-ray spectroscopy (EDX) in an JEOL JSM-6500 scanning electron microscope. The average stoichiometry was determined by EDX with multiple points examination on the crystals and the measured compositions are YbAlB$_{4}$ and Yb$_{0.81(2)}$Sr$_{0.19(3)}$AlB$_{4}$. Magnetization and electrical measurements were performed in Quantum Design MPMS-5XL and PPMS-9. Electric transport measurement results were obtained by a four-probe method using epoxy contacts. Sample dimensions were measured by an optical microscope Nikon SMZ-800 with 10 $\mu $m resolution. The geometries of the epoxy contacts for $\alpha -$YbAlB$_{4}$ and $\alpha -$Yb$_{0.81(2)}$Sr$_{0.19(3)}$AlB$_{4}$ are around 0.2 mm $\times$ 0.8 mm $\times$ 0.4 mm and 0.3 mm $\times$ 0.8mm $\times$ 0.4 mm, respectively. The uncertainty length is approximately 0.02 mm based on the contact shape in our optical microscope measurement. Therefore the relative error in geometry factor is at most 17.5\% for pure material and 14.5\% for Sr-doped crystal.

\section{Results and discussion}
\begin{figure}
\centerline{\includegraphics[scale=0.45]{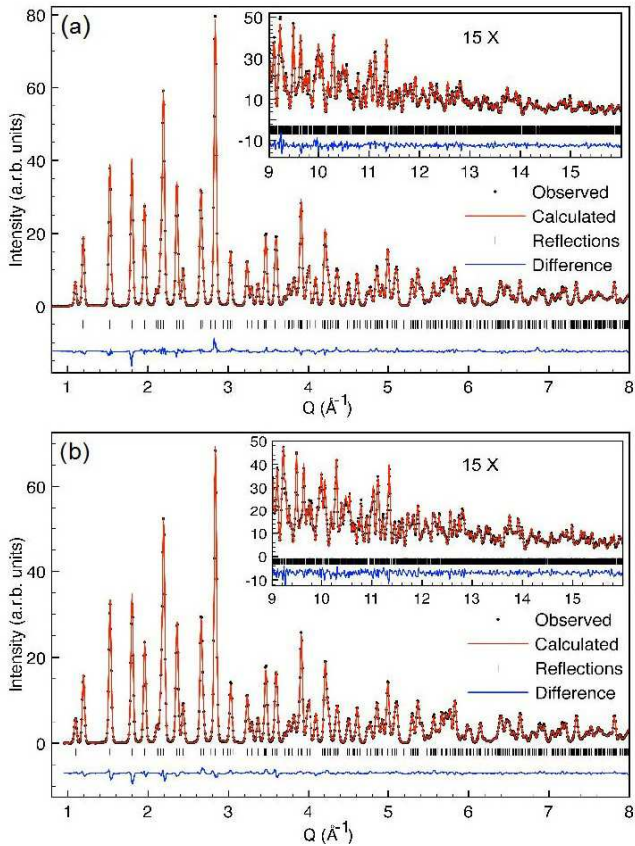}} \vspace*{-0.3cm}
\caption{(Color online) Rietveld refinements of the background subtracted synchrotron powder diffraction data of $\alpha$-YbAlB$_{4}$ (a) and $\alpha$-Yb$_{0.81(2)}$Sr$_{0.19(3)}$AlB$_{4}$ (b). Plots show the observed ($\bullet$) and calculated (solid red line) powder patterns with the difference curves (shown underneath each diffractogram) up to Q $\sim$16 ${\AA}$$^{-1}$. Vertical tick marks represent Bragg reflections in the $\alpha$-YbAlB$_{4}$ and $\alpha$-Yb$_{0.81(2)}$Sr$_{0.19(3)}$AlB$_{4}$ phases (space group Pbam). Fit residua, R(F**2), quantifying goodness of fit were 4\% for $\alpha$-YbAlB$_{4}$ and 3\% for $\alpha$-Yb$_{0.81(2)}$Sr$_{0.19(3)}$AlB$_{4}$.}
\end{figure}

The 2D XRD images were integrated into conventional 1D patterns with help of Fit2d computer program \cite{Hammersley}. Successful Rietveld analysis was carried out on both $\alpha$-YbAlB$_{4}$ and $\alpha$-Yb$_{0.81(2)}$Sr$_{0.19(3)}$AlB$_{4}$, using a single phase \textit{Pbam} structural model. Figure 1 shows model fits to the data. No impurity peaks were observed. Rietveld refinements produced excellent fits to the data up to a very high value of momentum transfer Q ($\sim$17 ${\AA}$$^{-1}$). This indicates the high purity of our samples and high quality of the XRD data. The Rietveld refinement was performed on XRD data using the General Structure Analysis System (GSAS/EXPGUI) \cite{Larson,Toby} computer package. A pseudo-Voigt function and a shifted Chebyshev polynomial were used to refine the peak profile and the background respectively. Gaussian and Lorentzian parameters \cite{Young} of the profile were refined after refining the zero-shift, lattice parameters, and the background. Then the atomic coordinates, occupation numbers, and the isotropic thermal displacement parameters (U$_{iso}$) were refined. At the last stage of the refinement, all profile and structural parameters were refined simultaneously to optimize the quality of fits and structural models. Atomic coordinates and U$_{iso}$'s agree well with  published values for $\alpha$-YbAlB$_{4}$ \cite{Macaluso}. Refined lattice parameters are a = 5.91821(12) ${\AA}$, b =  11.46239(22) ${\AA}$, and c =  3.49142(6) ${\AA}$ for $\alpha$-Yb$_{0.81(2)}$Sr$_{0.19(3)}$AlB$_{4}$ and a = 5.91878(12) ${\AA}$, b =  11.46491(23) ${\AA}$, and c =  3.49205(6) ${\AA}$ for $\alpha$-YbAlB$_{4}$. The unit cell volume of $\alpha$-Yb$_{0.81(2)}$Sr$_{0.19(3)}$AlB$_{4}$ (V=236.847(8) ${\AA}$$^{3}$) is 1.00(1) $\%$ smaller than in pure material. Yb(Sr) atomic positions in $\alpha -$YbAlB$_{4}$ have very high coordination number (CN = 14).\cite{Macaluso} Sr$^{2+}$ has larger radius (1.26 ${\AA}$) than Yb$^{3+}$ (0.98 ${\AA}$) and Yb$^{2+}$ (1.14 ${\AA}$) even at high CN = 8 where the data are available for comparison which suggests that small reduction in the lattice parameters could be attributed to the increase of metallic character of Sr-B bonds \cite{Shannon}.

\begin{figure}[tbp]
\centerline{\includegraphics[scale=0.7]{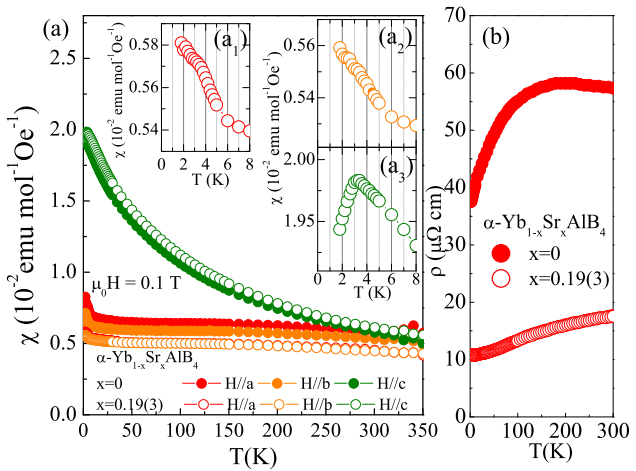}} \vspace*{-0.3cm}
\caption{(Color online) Magnetic susceptibility (a) and resistivity in the ab plane $\rho_{ab}$ (b) of $\alpha$-YbAlB$_{4}$ and $\alpha$-Yb$_{0.81(2)}$Sr$_{0.19(3)}$AlB$_{4}$. Insets in (a) present anisotropy in low temperature magnetic susceptibility of $\alpha$-Yb$_{0.81(2)}$Sr$_{0.19(3)}$AlB$_{4}$ around the magnetic transition.}
\end{figure}

Temperature-dependent magnetic susceptibility $\chi$=M/H taken in $\mu_{0}$H = 0.1 T shows Ising anisotropy where $\chi_{c}$ is strongly temperature dependent in contrast to $\chi_{a,b}$. This is consistent with previous reports (Fig. 2(a)) \cite{Macaluso,Matsumoto1}. Moreover, Fig. 2(a) shows small anisotropy in the $ab$-plane of $\alpha -$YbAlB$_{4}$. Similar to pure material, the $c$-axis magnetic susceptibility is also Curie-Weiss-like in $\alpha -$Yb$_{0.81(2)}$Sr$_{0.19(3)}$AlB$_{4}$, with no significant differences at high temperatures. As temperature decreases, $M/H$ for $\alpha -$Yb$_{0.81(2)}$Sr$_{0.19(3)}$AlB$_{4}$ $H$$\parallel$$b$-axis show weak anomalies at around 6 K ($H$$\parallel$$a,b$-axes) and 3 K ($H$$\parallel$$c$-axes). This implies complex and rather anisotropic magnetic ground state as $T\rightarrow0$. The easy axis of the antiferromagnetism is along the c-axis. Magnetic state and its anisotropy in $\alpha -$Yb$_{0.81(2)}$Sr$_{0.19(3)}$AlB$_{4}$ are probably different from the canted antiferromagnetic state in Fe doped $\alpha,\beta$-YbAlB$_{4}$ \cite{Kuga}. We note that magnetic ground state with small ordered moment ($\sim$ 0.6 $\mu_{B}$/Yb was predicted in $\alpha$-YbAlB$_{4}$ by the local density approximation
with on-site Coulomb repulsion correction (LDA+U) calculations, a testament to strong underlying magnetic correlations \cite{Nevidomskyy}. Due to the symmetry of Ce and Yb compounds that arises from the different electron configuration, $4f^{13}$ orbitals of Yb (``Kondo-hole'') can be considered as analog of $4f^{1}$ orbitals of Ce (``Kondo-electron''). Application of pressure favors smaller ionic radii. In contrast to Ce-based compounds where pressure delocalizes electrons on 4\textit{f} orbitals and promotes $4f^{1}$(J=5/2,Ce$^{3+}$)$\rightarrow$$4f^{0}$(J=0,Ce$^{4+}$) nonmagnetic ground state, in Yb-based compounds magnetism is induced with pressure since $4f^{14}$(J=0,Yb$^{2+}$)$\rightarrow$4f$^{13}$(J=7/2,Yb$^{3+}$). This has been observed in various Yb compounds \cite{Winkelmann2,Bauer}.

Effective moment was obtained from the high temperature part of the $c$-axis Curie-Weiss ($\chi =\chi _{0}+C/(T-\theta )$) fits. The $\chi _{0}$ is a temperature independent parameter that
includes the contribution of core diamagnetism, Pauli paramagnetism, and Landau diamagnetism. Curie constant C = N$_{A}$I$_{z}^{2}$/k$_{B}$, where N$_{A}$ is Avogadro number. This gives effective Ising moments $I_{z}$= 2.88(3)$\mu _{B}$/Yb for the $\alpha -$YbAlB$_{4}$ and $I_{z}$= 2.70(2)$\mu _{B}$/Yb for the $\alpha -$Yb$_{0.81(2)}$Sr$_{0.19(3)}$%
AlB$_{4}$ which is consistent with the previous reports for pure compound \cite{Macaluso,Matsumoto1} and implies no significant change with Sr substitution. The Curie-Weiss temperatures are -150(4) K and -152(6) K for $\alpha -$YbAlB$_{4}$ and $\alpha -$Yb$_{0.81(2)}$Sr$_{0.19(3)}$AlB$_{4}$, respectively. We note that Curie-Weiss temperatures in literature vary from -110(5) K to -190(9) K \cite{Macaluso,Kuga}. The discrepancies could be due to the fitting range, presence of surface impurities but also to disorder as we discuss below.

Temperature dependent resistivity $\rho_{xx}(T)$ of $\alpha -$YbAlB$_{4}$ (Fig. 2(b)) is consistent with previous result \cite{Macaluso,Matsumoto1}, however the residual resistivity value is larger when compared to crystals used in previous studies \cite{Matsumoto1}. We note that in our samples of $\alpha -$YbAlB$_{4}$ resistivity values varied by the factor of 3 at room temperature, much larger than the uncertainty in sample geometry. This attests to considerable contribution of the crystalline disorder scattering (the sample with lowest $\rho_{0}$ is shown in Fig. 2(b)). The approximate residual resisitivity of $\alpha -$YbAlB$_{4}$ and $\alpha -$Yb$_{0.81(2)}$Sr$_{0.19(3)}$AlB$_{4}$ is 36 $\mu\Omega$cm and 11 $\mu\Omega$cm, at 2 K respectively. Strontium substitution lowers the residual resistivity, however both crystals show metallic behavior. A broad peak in $\rho(T)$ of our disordered $\alpha -$YbAlB$_{4}$ crystals at about 200 K corresponds to $\sim$ 250 K $\rho(T)$ peak of crystals used in Ref. 13. This suggests that disorder in $\alpha -$YbAlB$_{4}$ is intimately connected with magnetic disorder, i.e. with Yb$^{3+}$ ions since the coherence peak in our crystals is observable without subtraction of $\rho$ of nonmagnetic analog $\alpha$-LuAlB$_{4}$.

\begin{figure}[tbp]
\centerline{\includegraphics[scale=0.7]{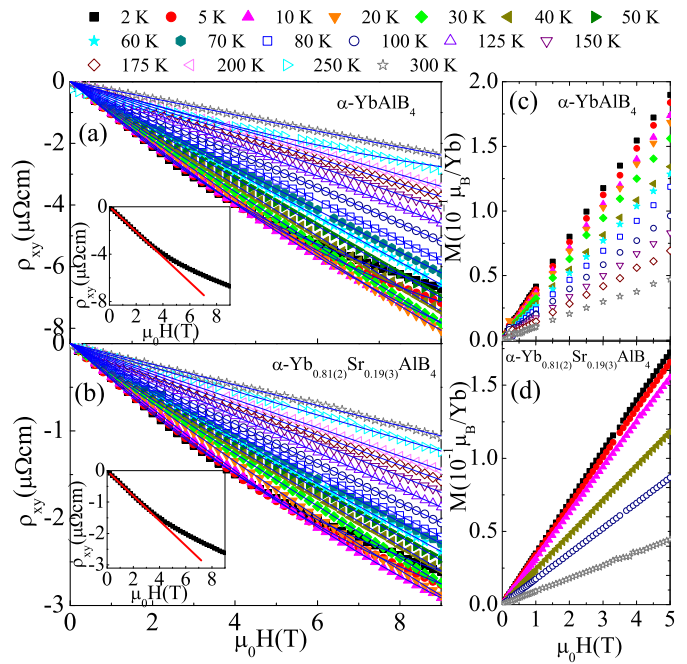}} \vspace*{-0.3cm}
\caption{(Color online) Hall resistivity $\rho_{xy}$ (a,b) and corresponding magnetization for $\mu_{0}$H $\parallel$c-axis (c,d) of $\alpha$-YbAlB$_{4}$ and $\alpha$-Yb$_{0.81(2)}$Sr$_{0.19(3)}$AlB$_{4}$. Solid lines (a,b) represent fit results (see text). Inset shows $\rho_{xy}$ at 2 K for clarity. }
\end{figure}

Magnetic field dependence of Hall resistivity ($\rho _{xy}$) for several
different temperatures from (2 -300) K are shown in Fig. 3 (a) and (b) for $\alpha -$YbAlB$_{4}$ and $\alpha -$Yb$_{0.81(2)}$Sr$_{0.19(3)}$AlB$_{4}$, respectively. The same sample was used for both $\rho_{xx}(T)$ and $\rho_{xy}(H)$ measurements. The $\rho _{xy}(H)$ is linear in the magnetic field at high temperatures, suggesting an ordinary Hall effect [$\rho _{xy}/B = R_{H} = -1/ne$] and conduction in a single band model with a carrier concentration $n$. However, when T$\leq$10 K $\rho _{xy}(H)$ is curved (insets in Fig. 3(a,b)), similar to $\beta -$YbAlB$_{4}$ \cite{OFarrel}. This is reminiscent of the anomalous Hall effect \cite{OHandley}: $\rho _{xy}(H)=R_{0}H+R_{s}M(H),$ where $R_{0}$ and $R_{s}$ are the normal and spontaneous Hall constants, and $M$ is sample magnetization. Attempts to fit $\rho _{xy}(H)$ (Fig. 3(a,b)) using $M(H)$ obtained at the same temperature (Fig. 3(c,d)) were unsuccessful. Hence, the anomalous Hall effect is unlikely cause of the $\rho _{xy}(H)$ nonlinearity. We now proceed with the two-band analysis of electronic transport for $\rho _{xy}(H)$ for T $\leq$ 10 K. Hall resistivity in the two band model is:


\begin{eqnarray}
&\rho _{xy}&/\mu _{0}H = R_{H}= \nonumber \\
&&\frac{1}{e}\frac{(\mu _{h}^{2}n_{h}-\mu _{e}^{2}n_{e})+(\mu _{h}\mu
_{e})^{2}(\mu _{0}H)^{2}(n_{h}-n_{e})}{(\mu _{e}n_{h}+\mu
_{h}n_{e})^{2}+(\mu _{h}\mu _{e})^{2}(\mu _{0}H)^{2}(n_{h}-n_{e})^{2}}
\end{eqnarray}

where $n_{e}$, $n_{h}$, $\mu _{e}$, and $\mu _{h}$ are carrier density and mobility of electron and hall bands, respectively \cite{Smith}. Fits to the single band model (high temperatures) and two band model (T $\leq$ 10K) are excellent. Interestingly, the multiband transport becomes important for T$\leq$10 K, which is close to $T^{*}$ temperature scale that marks the onset of a heavy fermion Fermi liquid \cite{Matsumoto1}.

\begin{figure}[tbp]
\centerline{\includegraphics[scale=0.7]{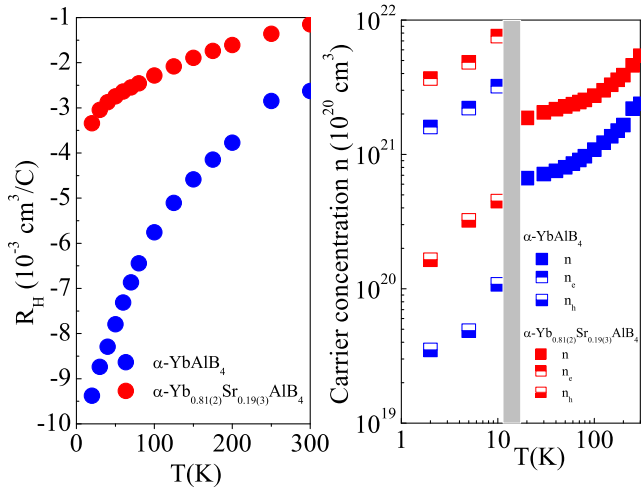}} \vspace*{-0.3cm}
\caption{(Color online) Hall constant $\rho _{xy}/B = R_{H} = -1/ne$ (a) and carrier concentrations for $\alpha -$YbAlB$_{4}$ and $\alpha -$Yb$_{0.81(2)}$Sr$_{0.19(3)}$AlB$_{4}$. Shaded area marks the crossover from single band ($T\leq10K$) to two band conduction ($T\geq20K$).}
\end{figure}

As opposed to $\beta -$YbAlB$_{4}$ \cite{OFarrel}, we do not observe a peak of $R_{H}$ at 40 K (Fig. 4(a)). The peak might appear at a similar temperature in the case of $\alpha -$YbAlB$_{4}$ at its clean limit. The difference might come from the elastic scattering that generates the residual resistivity, which seems to substantially change the temperature dependence of the longitudinal resistivity. $R_{H}$ (Fig. 4(a)) is negative over the measured temperature range and its magnitude decreases in $\alpha -$Yb$_{0.81(2)}$Sr$_{0.19(3)}$AlB$_{4}$ whereas the carrier concentration $n$ increases (Fig. 4(b), consistent with the resistivity result (Fig. 1(b)) since Sr substitution promotes metallic character of bonds. Since both $\alpha -$YbAlB$_{4}$ and $\beta -$YbAlB$_{4}$ are metallic and since structural studies have confirmed nanoscale phase separation in both polymorphs (\textit{Cmmm} and \textit{Pbam} space group coexistence in both crystals) \cite{YubutaK}, Hall constant might be affected by sample differences.

Mobility values extracted from (1) for T$\leq$10 K where the multiband transport is dominant are shown in Tables 1 and 2. It can be observed that formation of heavy fermion Fermi liquid probably involves
both electron and hole bands whereas hole bands have somewhat larger Kondo coupling. Moreover, Sr$^{2+}$ substitution on Yb atomic sites does not induce significant change in mobility values. We note that electron and hole Fermi surface parts were also found in $\beta -$YbAlB$_{4}$ \cite{Ramires}.

\begin{table}[tbp]\centering%
\caption{Summary of mobility values for $\alpha -$YbAlB$_{4}$ in multiband regime for T$\leq$ 10 K}%
\begin{tabular}{ccc}
\hline\hline\hline
$\alpha -$YbAlB$_{4}$ & $\mu_{e}$ (m$^{2}$/Vs) & $\mu_{h}$ (m$^{2}$/Vs)\\
2 K & 0.12(1) & 0.07(1)\\
5 K & 0.09(1) & 0.04(1)\\
10 K & 0.06(1) & 0.02(0)\\
\hline\hline
\end{tabular}%
\label{4}%
\end{table}%

\begin{table}[tbp]\centering%
\caption{Summary of mobility values for $\alpha -$Yb$_{0.81(2)}$Sr$_{0.19(3)}$AlB$_{4}$ in multiband regime for T$\leq$ 10 K}%
\begin{tabular}{ccc}
\hline\hline\hline
$\alpha -$Yb$_{0.81(2)}$Sr$_{0.19(3)}$AlB$_{4}$ & $\mu_{e}$ (m$^{2}$/Vs) & $\mu_{h}$ (m$^{2}$/Vs)\\
2 K & 0.13(0) & 0.08(1)\\
5 K & 0.10(1) & 0.05(1)\\
10 K & 0.06(2) & 0.03(1)\\
\hline\hline
\end{tabular}%
\label{4}%
\end{table}%

\section{Conclusion}

In summary, we showed that 19(3) \% of Sr substitution on Yb site in $\alpha -$YbAlB$_{4}$ induces magnetic ground state below 3 K. Both $\alpha -$YbAlB$_{4}$ and $\alpha -$Yb$_{0.81(2)}$Sr$_{0.19(3)}$AlB$_{4}$ show signature of multiband electronic transport for T$\leq$ 10 K. Increased conductivity of $\alpha -$Yb$_{0.81(2)}$Sr$_{0.19(3)}$AlB$_{4}$ could be attributed to slight lattice contraction with Sr doping that promotes Yb$^{3+}$ state, i.e. enhancement of electron density by hybridization of $4f$ orbital and conduction electron bands. This promotes magnetic ground state via the RKKY mechanism since T$_{RKKY}$ $\sim$ k$_{F}$$r$ where $k_{F}$ is the Fermi wavevector and $r$ is the Yb$^{3+}$ distance. The formation of heavy fermion Fermi liquid takes place in both electron and hole bands whereas the quasiparticle mass of heavy fermion bands is unchanged in Sr substituted crystals. Further band structure and experimental studies are needed to fully explain why $\alpha -$YbAlB$_{4}$ and $\beta -$YbAlB$_{4}$ have profoundly different ground states despite similar crystal structure and multiband electronic transport.

\section*{Acknowledgments}
Work at Brookhaven is supported by the U.S. DOE under Contract No. DE-AC02-98CH10886. This work has benefited from using the X7B beamline of the NSLS at Brookhaven Laboratory. We thank John B. Warren for help with scanning electron microscopy measurements, and Jonathan Hanson for help with x-ray measurements. This work is partially supported by Grant-in-Aid for Scientific Research (No. 25707030) from JSPS, Japan. C. P. acknowledges ISSP at the University of Tokyo for its hospitality and financial support.

\dag Present address: Advanced Light Source, E. O. Lawrence Berkeley
National Laboratory, Berkeley, California 94720, USA.
\ddag petrovic@bnl.gov

\end{document}